# Astro2020 Science White Paper

# Planetary Science with Astrophysical Assets:
## *Defining the Core Capabilities of Platforms*

**Thematic Areas:** ☒ Planetary Systems  ☒ Star and Planet Formation

**Principal Author:**
Name: James M. Bauer (Co-Chair)
Institution: University of Maryland
Email: gerbsb@astro.umd.edu
Phone: (301) 405-6218

**Co-authors:**
*Stefanie Milam (Co-Chair; GSFC) - Stefanie.n.milam@nasa.gov*
*Gordon Bjoraker (GSFC) - gordon.l.bjoraker@nasa.gov*
*Sean Carey (IPAC) - sean.j.carey-117106@jpl.nasa.gov*
*Doris Daou (NASA HQ) - doris.daou-1@nasa.gov*
*Leigh Fletcher (University of Leicester) - leigh.fletcher@leicester.ac.uk*
*Walt Harris (University of Arizona) - wharris@lpl.arizona.edu*
*Paul Hartogh(MPI - SSRG) - hartogh@mps.mpg.de*
*Christine Hartzell (UMD) - hartzell@umd.edu*
*Amanda Hendrix (PSI) - ahendrix@psi.edu*
*Carrie Nugent (Olin College) - nugent.carolyn@gmail.com*
*Andy Rivkin (APL, JHU) - Andy.Rivkin@jhuapl.edu*
*Timothy Swindle (LPL, U of AZ) - tswidle@lpl.arizona.edu*
*Cristina Thomas (NAU) - cristina.thomas@nau.edu*
*Matthew S. Tiscareno, SETI Institute - matt@seti.org*
*Geronimo Villanueva (GSFC) - geronimo.villanueva@nasa.gov*
*Scott Wolk (CfA, Harvard) - swolk@cfa.harvard.edu*

**Abstract**: We seek to compile a uniform set of basic capabilities and needs to maximize the yield of Solar System science with future Astrophysics assets while allowing those assets to achieve their Astrophysics priorities. Within considerations of cost and complexity, inclusion of capabilities that make a particular platform useable to planetary science provide a critical advantage over platforms lacking such capabilities.

# Planetary Science with Astrophysical Assets:
## *Defining the Core Capabilities of Platforms*

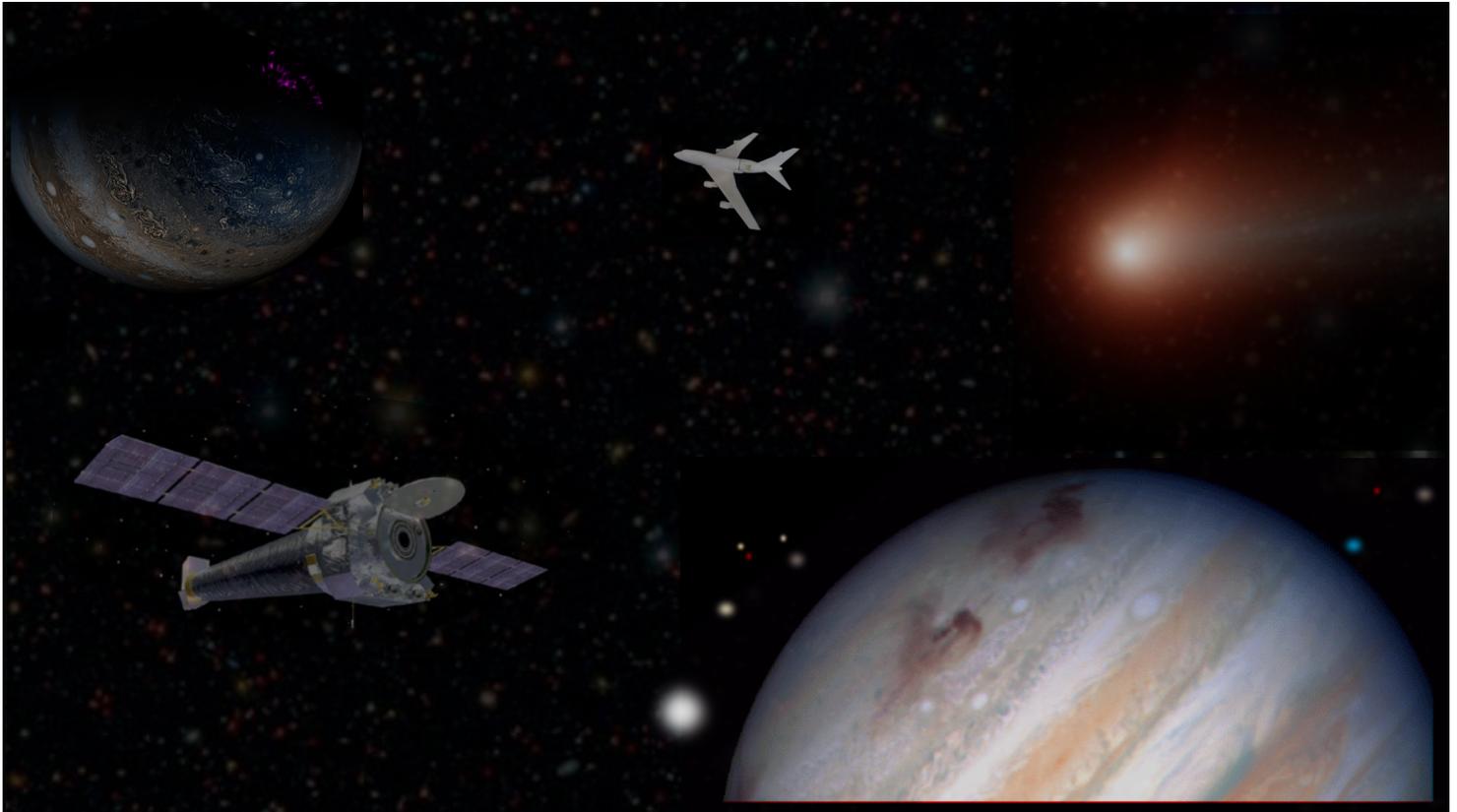

Small Bodies Assessment Group Committee on Planetary Science with Astrophysical Assets

*James Bauer (Co-Chair; gerbsb@umd.edu), Stefanie Milam (Co-Chair; GSFC), Gordon Bjoraker (GSFC), Sean Carey (IPAC), Doris Daou (NASA HQ), Leigh Fletcher (University of Leicester), Walt Harris (University of Arizona), Paul Hartogh (MPI - SSRG), Christine Hartzell (UMD), Amanda Hendrix (PSI), Carrie Nugent (Olin College), Andy Rivkin (APL, JHU), Timothy Swindle (LPL, U of AZ), Matthew Tiscareno (SETI), Cristina Thomas (NAU), Geronimo Villanueva (GSFC), Scott Wolk (CfA, Harvard)*

**Thematic Areas:** Planetary Systems, Star and Planet Formation

**Abstract**: We seek to compile a uniform set of basic capabilities and needs to maximize the yield of Solar System science with future Astrophysics assets while allowing those assets to achieve their Astrophysics priorities. Within considerations of cost and complexity, inclusion of capabilities that make a particular platform useable to planetary science provide a critical advantage over platforms lacking such capabilities.

## I. Introduction

Planetary science is the study of Solar System bodies, objects which in principle can be visited by spacecraft within a span of years. Even the most distant of populations generally have interlopers that visit the accessible reaches of the Solar System (cf. Oort 1950, Meech et al. 2018). Yet for Astrophysics, which generally includes the study of bodies outside our Solar System, even the nearest bodies are inaccessible to visits by spacecraft on practical timescales, and so must be studied exclusively by telescopes. Hence when we speak of astrophysical assets, we mean, almost exclusively, telescopes, with some notable exceptions. Astrophysical observations are obtained almost exclusively through the use of telescopes, but the use of telescopes is certainly not exclusively astrophysical, and here is where a need arises for the delineation of capabilities that facilitate wide utility of these assets to the study of planetary science. This is because, for the present, the study of large numbers of Solar System bodies by spacecraft visits, and emplacement of spacecraft simultaneously with time-critical events, are similarly as impractical as sending spacecraft to even the nearest exoplanetary systems.

We present a uniform set of basic capabilities and needs to maximize the yield of Solar System science with future Astrophysics assets while allowing those missions to achieve their Astrophysics priorities. The capabilities we discuss are prioritized and presented in sub-categories and explained in relation to major sub-disciplines within the field of planetary science. Table 1 lists particular sub-disciplines and major capabilities and science they facilitate.

## II. Non-sidereal Tracking

The term planet comes from the greek "wanderer", so called because when they are viewed, Solar System objects move relative to the background stars. Hence it is widely regarded that the single most important capability in facilitating planetary science with telescopic assets is their ability to track at specified non-sidereal rates. Most telescopes have the ability to slew, stabilize, and rectify tracking on stellar sources within their regular operational and scheduling constraints regardless of their own motion through space, and so most have the mechanical controls and guidance hardware to enable non-sidereal tracking (NST), though software implementation and additional modest scheduling overheads may be necessary in application.

Telescopes must design within the bounds of their trade-spaces to accommodate non-sidereal tracking capabilities. Spitzer Space Telescope, for example, was capable of guiding at a considerable non-sidereal rate of 1 arcsecond per second but was unable to vary its tracking rate in mid-observation. Such limitations did not seem to significantly impact even near-earth object (NEO) survey programs with the spacecraft (cf. Trilling et al. 2010). The James Webb Space Telescope (JWST), on the other hand, has set its nominal limits to a smaller rate of 30 milli-arcseconds per second (Milam et al. 2016), but with a reasonable justification: (1) that this limit accommodated the non-sidereal motion of Mars, the only terrestrial planet within its field of regard, (2) that the telescope could also guide with differential tracking rates throughout an observation, and (3) that the limit avoids significant degradation of the JWST's imaging capabilities afforded by its 6.5 meter aperture. However, failure to accommodate non-sidereal tracking altogether greatly imperils the ability of a platform to have a fully active planetary science program and limits the science that can be otherwise achieved with the platform across all the major planetary science sub-disciplines by impacting the angular resolution, sensitivity, and astrometric and photometric accuracy. This is especially true for telescopes with targeted observation capabilities, since comets, asteroids, and giant and terrestrial planets all have significant non-sidereal motions, typically ranging into the 10s to 100s of arcseconds/hour.

### III. Field of View / Field of Regard

*Field of View (FOV):* Large fields of view are of use to planetary and astrophysical surveys alike. Special modes of large-array surveys, however, coupled with rapid-read-out capabilities or non-sidereal tracking, aid planetary observations. Particular categories of small bodies may be targeted for discovery by tailoring non-sidereal tracking rates to their median motions appropriate to their viewing geometries at the time of observation. Alternatively, for characterization purposes, series of images with brief exposure times rapidly read-out coupled with low-noise from detectors may mitigate variations in motion for a range of planetary objects across the large fields of view.

*Field of Regard (FOR)*: Limitations of the field of regard, the range of elongations that may be viewed by a telescope, especially impact planetary science observations of associated time-domain events. Such events, including naturally occurring events such as mutual events (Grundy et al. 2012), ring-plane crossings (Scharringhausen and Nicholson 2013), and planetary impacts (Hammel et al. 1995), as well as spacecraft encounters with planetary bodies (cf. Lisse et al. 2006), play an indispensable role in the characterization of Solar System objects. Extremely narrow ranges of the FOR limit the usage of a telescope to serendipitous observations of planetary targets, i.e. whether they are crossing within the FOR of the telescope at the desired time of their orbit. While FOR considerations will always fall within the trade space of a telescope's design and cost, and designing shielding that allows for a larger range of accessible elongations can be significant, even marginal extension of the FOR may open access to high-priority time critical events.

### IV. Scheduling and MOS-GDS

*Adequate single-frame and meta-data:* Owing to the temporal nature and proximities of Solar System objects, the disparity between the analysis and housekeeping data for astrophysical and planetary observations is often overlooked. However, the basic elements that facilitate planetary observations are often already included or easily accommodated through the mission operations and ground based analysis systems (MOS/GDS).

Owing to the motion of planetary objects on the sky, it is essential that individual observations be kept. In other words, for telescopes, ground data systems must archive the individual images. Some information can be obtained through the records of individual detections with accurate astrometry (preferably sub-arcsecond) and timestamps (preferably to better than a few tenths of a second). However, for context and photometric information, including identification of extended morphology, nothing is as informative as the image itself, with the accompanying meta-data often provided in the header.

In order to facilitate sufficient astrometry, accurate station keeping is necessary. For many Solar System objects (e.g. more distant populations), this is not a significant constraint. Objects at distances of 1 AU or more from the platform can obtain sub-arcsecond astrometric accuracy with uncertainties of a few km. However, for NEOs (comets or asteroids), station-keeping with positional uncertainties of less than 1km are necessary for sub-arcsecond astrometry for objects approaching within half a lunar distance.

*Flexible Scheduling to accommodate planetary observations:* Mission operations can play a key role in the accommodation of planetary observations by implementing short-term target-of-opportunity programs for moving objects. These should not have penalties that are disproportionate with respect to inertial targets. Furthermore, tools to plan and implement such TOO planetary observations should be considered in lock-step with the astrophysical observation tools. Such scheduling is critical to observing unanticipated phenomena, such as the discovery and characterization of small bodies on hyperbolic or planet-impacting orbits, impact events, the sudden appearance of storms on giant planets or Mars, close-encounters of comets with terrestrial planets, active Centaurs or disrupting asteroids, to name a few. Similarly, mid-term (e.g. weeks) or long-term (years) monitoring of cometary small bodies or changing planetary atmospheric activity are valuable capabilities for planetary science that should be considered.

## V. Imaging Arrays, Dynamic Range and Read-Out

Science yielded through imaging of planetary bodies benefits from many of the same capabilities as astrophysical objects. Higher angular resolution and unprecedented photometric sensitivity across wavelengths provide detail at smaller physical scales and accurate fluxes from faint sources. However, telescopic studies of planetary objects also benefit from less commonly utilized capabilities regarding the imaging instrument design. Observations of most major categories of objects, from small bodies like comets, volcanically active satellites, terrestrial, and giant planets alike can make use of large dynamic range coupled with increased sensitivities. Modes of rapid read-out, at rates of several hertz, for example, can allow for averaging over co-adds with brief enough exposures to avoid most instances of saturation. Rapid sampling modes, including sub-array read-out capabilities, can also facilitate observing of planetary occultations and mutual events, employing time-domain observing methods through which many key planetary science discoveries have been made. Imaging arrays, which are utilized in both spectroscopic and filtered-band-pass imaging, have increased application to planetary time-domain astronomy with these capabilities.

## VI. Spectroscopy: Wavelength Range and Resolution, and IFS's

As with imaging, planetary observations profit from improvements in spectral resolution and sensitivity, as well as the ability to sample spectra uniformly over an imaging plane, as with an integral field spectrometer (IFS). The costs and benefits of these capabilities are dictated primarily by the astrophysical asset's science scope, but some flexibility to marginally expand critical wavelength range or access at the desired and scoped resolution capabilities should receive consideration when such modifications encompass key spectral features, like key emission or absorption lines in cometary or planetary atmospheres, or asteroidal mineralogical features. Such features that are only accessible to specific platforms, such as $CO_2$ emission lines in comets that are only accessible from space, should be considered especially if the measurement capability does not greatly alter the Astrophysics science goals or platform costs.

## VII. Surveys and Cadence

Efficiency is a major consideration in the trade space of surveys. Discoveries of time-domain events often require multiple observations, and so often facilitate the multiple observations necessary to establish orbits of newly discovered bodies and possibly constrain elements of their rotation light curves. However, surveys conducted with the primary intent of opening new wavelength regimes at significantly higher resolution, detail, or sensitivity may be less effective at the discovery of Solar System objects unless the objects are detected multiple times over appropriately spaced time intervals. Cadences on the order of hours or tens of minutes are sensitive to non-sidereal motions out to Centaur and Kuiper Belt distances, for example, and effective surveys may obtain sufficient astrometric measurements (e.g. 4 or more observations over a 2-12 hour period and multiple visits over several days or a few weeks) if they are planned to include a component of Solar System objects discovery. The appropriate cadence must be considered for the object class (usually a subcategory of small body) that is intended to be discovered and identified within the survey data. For surveys with only a few observations per night, or a only a few observations per month, a single extra exposure or visit per area within the survey cadence can significantly boost the rate of discovery, detection, or characterization.

## VIII.    Summary

We have outlined the capabilities that are of general interest to implementing planetary science programs using observing platforms among astrophysical assets. Such capabilities should be considered early in the mission design process. The committee's conclusions were not intended to be comprehensive, but rather to serve as a general guide. Prioritization of the capabilities and a general sense of their values to the associated sub-categories of planetary objects is also provided within the table. These capabilities are independent of wavelength, in general. Other capabilities may be of special interest to a particular sub-discipline of planetary science should be considered in those contexts.  However, regarding the trade-spaces of cost and complexity, the inclusion of capabilities that make a particular platform useable to planetary science provide a critical advantage over platforms lacking such capabilities, and often facilitate particular benefits to Astrophysical science as well.

**Enhanced Capabilities and Benefits**

| Capability/Priority | Small Bodies | Giant &Terrestrial Planets |
|---|---|---|
| *NST (1)* | • Allows full benefits of the observing platform to be used for planetary bodies. | |
| *FOV/FOR (2)* | • Large FOV allows simultaneous imaging of cometary coma, tails, and trails.<br>• Expanded FOR allows access of small bodies through more of their orbits. | • A small FOV can impact the ability to monitor the planet's full disk, or conduct simultaneous satellite observations.<br>• A limited FOR impacts the ability to study a given planet in the time-domain. |
| *Scheduling, MOS-GDS (3)* | • Flexibility in scheduling without special penalties for planetary targets allows more complete exploration of the time-domain.<br>• Complete meta-data and well calibrated individual frames facilitates small body discoveries and characterization of temporal behavior. | |
| *Imaging, Dynamic Range and Readout (4)* | • Larger dynamical range allows for comprehensive comet imaging.<br>• Planetary occultation observations are facilitated by rapid read-out. | • Limited dynamic range impacts imaging and spectroscopy of atmospheres, aurorae, and satellite volcanism. |
| *Spectroscopy (5)* | • Flexibility for marginal alterations of wavelength coverage to include key spectral features can greatly impact planetary spectroscopy across categories. | |
| *Survey cadence (6)* | • For surveys, appropriate cadence considerations tailored to the planetary bodies of highest interest are critical to the survey success. | |